\documentclass{emulateapj}
\usepackage{times}

\newcommand{\XMM}{{\em XMM-Newton }}
\newcommand{\Ch}{{\em Chandra }}

\def\gappeq{\mathrel{ \rlap{\raise.5ex\hbox{$>$}}
                      {\lower.5ex\hbox{$\sim$}}  } }
\def\lappeq{\mathrel{ \rlap{\raise.5ex\hbox{$<$}}
                      {\lower.5ex\hbox{$\sim$}}  } }

\shorttitle{XMM-Newton Observations of 3C\,305, DA\,240, and 4C\,73.08}
\shortauthors{EVANS ET AL.}

\begin{document}

\title{XMM-Newton Observations of the Nuclei of the Radio Galaxies 3C\,305, DA\,240, and 4C\,73.08}
\author{Daniel~A.~Evans\altaffilmark{1,2}, Martin~J.~Hardcastle\altaffilmark{3}, Julia~C.~Lee\altaffilmark{1,2}, Ralph~P.~Kraft\altaffilmark{2}, Diana~M.~Worrall\altaffilmark{4}, Mark~Birkinshaw\altaffilmark{4}, Judith~H.Croston\altaffilmark{3}}
\altaffiltext{1}{Harvard University, Department of Astronomy, 60 Garden Street, Cambridge, MA 02138}
\altaffiltext{2}{Harvard-Smithsonian Center for Astrophysics, 60 Garden Street, Cambridge, MA 02138}
\altaffiltext{3}{School of Physics, Astronomy \& Mathematics, University of Hertfordshire, College Lane, Hatfield AL10 9AB, UK}
\altaffiltext{4}{University of Bristol, Department of Physics, Tyndall Avenue, Bristol BS8 1TL, UK}

\begin{abstract}

We present new \XMM EPIC observations of the nuclei of the nearby radio galaxies 3C\,305, DA\,240, and 4C\,73.08, and investigate the origin of their nuclear X-ray emission. The nuclei of the three sources appear to have different relative contributions of accretion- and jet-related X-ray emission, as expected based on earlier work. The X-ray spectrum of the FRII narrow-line radio galaxy (NLRG) 4C\,73.08 is modeled with the sum of a heavily absorbed power law that we interpret to be associated with a luminous accretion disk and circumnuclear obscuring structure, and an unabsorbed power law that originates in an unresolved jet. This behavior is consistent with other narrow-line radio galaxies. The X-ray emission of the low-excitation FRII radio galaxy DA\,240 is best modeled as an unabsorbed power law that we associate with a parsec-scale jet, similar to other low-excitation sources that we have studied previously. However, the X-ray nucleus of the narrow-line radio galaxy 3C\,305 shows no evidence for the heavily absorbed X-ray emission that has been found in other NLRGs. It is possible that the nuclear optical spectrum in 3C\,305 is intrinsically weak-lined, with the strong emission arising from extended regions that indicate the presence of jet--environment interactions. Our observations of 3C\,305 suggest that this source is more closely related to other weak-lined radio galaxies. This ambiguity could extend to other sources currently classified as NLRGs. We also present \XMM and VLA observations of the hotspot of DA\,240, arguing that this is another detection of X-ray synchrotron emission from a low-luminosity hotspot.

\end{abstract}


\section{INTRODUCTION}
\label{intro}

Radio galaxies consist of twin jets of particles that are ejected from a compact region in the vicinity of a supermassive black hole, feeding into large-scale `plumes' or `lobes'. There are two principal morphological classes of radio galaxies, low-power (Fanaroff-Riley type I, hereafter FRI) sources and high-power (FRII) sources \cite{fr74}. FRI sources exhibit `edge-darkened' large-scale radio structure, and modeling implies that initially supersonic jets in these sources decelerate to transonic speeds on $\sim$kpc scales before flaring into large plumes (e.g., \citealt{per07}). FRII sources appear `edge-brightened', and in these cases highly supersonic jets propagate out to large distances (often $>100$ kpc) from the core before terminating in bright hotspots and accompanying radio lobes. Observationally, the Fanaroff-Riley divide occurs at a 178-MHz radio power of $\sim$$10^{25}$ W~Hz$^{-1}$~sr$^{-1}$.

It is important to understand whether the kpc-scale Fanaroff-Riley
dichotomy is determined by the interaction between the jet and its external
hot-gas environment (e.g.,
\citealt{bic95}), or rather is nuclear in origin and governed by
differences in the properties of the accretion flow (\citealt{rey96}).
The first observations of large samples of $z<0.1$ 3CRR radio-galaxy {\it nuclei} with \Ch and \XMM (\citealt{don04,bal06,evans06}) showed that FRI nuclei show no signs of heavily absorbed X-ray emission that would be expected from standard AGN unification models (\citealt{up95}), are dominated by emission from an unresolved jet (e.g.,~\citealt{wb94,bal06,evans06}), and have highly radiatively inefficient accretion flows. Narrow optical-line FRII sources show evidence for heavily obscured ($N_{\rm H}>10^{23}$~cm$^{-2}$) nuclear X-ray emission that is associated with a radiatively efficient accretion flow, together with an unabsorbed component of jet-related emission (\citealt{evans06,hec06}). FRII radio galaxies at higher redshift are consistent with such behavior (\citealt{bel06}).

A significant breakthrough for understanding the physical origin of
the FRI/FRII dichotomy came from \Ch and \XMM observations of the
population of {\it low-excitation radio galaxies} (LERGs), which have weak or no
emission lines in their optical spectra (\citealt{hine79,jac97}).
Almost all FRI radio galaxies are LERGs, but there is a significant
population of FRII LERGs at $0.1 < z < 0.5$. The X-ray spectra of
LERGs, irrespective of their FRI or FRII morphology, are dominated by
unabsorbed emission that can be associated with a parsec-scale jet, with no obvious contribution from
accretion-related emission. These sources are likely to accrete in a radiatively
inefficient manner (\citealt{hec06}). On the other hand,
high-excitation radio galaxies (HERGs -- i.e., NLRGs, BLRGs, and
quasars), which display prominent narrow or broad optical emission
lines, have X-ray spectra that are consistent with standard
unification models: they show evidence for luminous, radiatively
efficient accretion disks, together with circumnuclear tori when the
source is oriented close to edge-on with respect to the observer. HERGs tend to show evidence for additional hot dust over and above that of LERGs in their mid-IR spectra (e.g., \citealt{ogle06}; Birkinshaw et al., in preparation), which is again consistent with reprocessing of luminous accretion-related emission by torus-like structure. Most
high radio-power (FRII) sources are high-excitation radio galaxies.
The distinct X-ray nuclear properties of low- and high-excitation radio galaxies, regardless of their large-scale FRI or FRII morphology, could be interpreted as implying that the Fanaroff-Riley dichotomy remains principally influenced by jet power and environment. The excitation dichotomy, on the other hand, is interpreted to be attributed to the radiative efficiency of the accretion flow (e.g., \citealt{hec06}) and possibly related to the nature of the accreting material (\citealt{hec07}).


Here, we report new \XMM observations of the nuclei of three
$z<0.1$ 3CRR radio galaxies --- 3C\,305, DA\,240, and
4C\,73.08. The three sources have 178-MHz radio powers that lie close to the FRI/FRII dividing luminosity (Table~\ref{sourcessummary}), plus a range of radio morphologies and optical emission-line characteristics. They are therefore good candidates for examining possible connections between the central engine and large-scale radio characteristics. This paper is organized as follows. In Section 2, we describe the optical and radio properties of the three sources. Section 3 contains a description of the data and a summary of our analysis. In Section 4, we report the results of our spectroscopic analysis of the sources. In Section 5, we describe VLA and \XMM observations of the bright NE hotspot in DA\,240. In Section 6, we interpret the observations in the context of our previous \Ch and \XMM observations of 3CRR radio galaxies and discuss the optical emission-line characteristics of the sources. We end with our conclusions in Section 7. All results presented in this paper use a cosmology in which $\Omega_{\rm m, 0}$ = 0.3, $\Omega_{\rm \Lambda, 0}$ = 0.7, and H$_0$ = 70 km s$^{-1}$ Mpc$^{-1}$. Errors quoted in this paper are 90 per cent confidence for one parameter of interest (i.e., $\chi^2_{\rm min}$ + 2.7), unless otherwise stated.

\section{Overview Of The Sources}

\subsection{3C\,305}

3C\,305 is a $z=0.0416$ ($d_{\rm L}=183$~Mpc) narrow-line radio galaxy
(\citealt{lrl83}) with an unusually compact radio morphology that
displays both FRI and FRII characteristics. MERLIN and VLA
observations of the source (\citealt{hec82,jac03,mor05}) show twin
jets that each extend into radio lobes separated by $\sim$4$''$
(3.3~kpc). \cite{lrl83} classify the source
as an FRI-type radio galaxy. Optical emission-line studies of the
circumnuclear environment of 3C\,305 show an extended morphology
(\citealt{hec82,jac95}), and a detailed comparison between the radio
ejecta and [O~{\sc ii}] gas observed with {\it HST} (\citealt{jac95})
suggests that the gas has been shock-excited by the jet.

\subsection{DA\,240}

DA\,240 ($z=0.0356$, $d_{\rm L}=157$~Mpc) is a giant radio galaxy (GRG), the
name given to the
subclass of FRII sources with a projected radio extent in excess of
1~Mpc. Westerbork Synthesis Radio Telescope (WSRT) images of the
source (\citealt{klein94,peng04}) show two hotspots, with the northeastern
one 50 times brighter than the southwestern one. The northeastern hotspot has an
unusual bifurcated structure, with what appears to be a radio jet
entering a compact, primary hotspot, together with a fainter secondary hotspot feature. Optically, DA\,240 is classified as a low-excitation radio galaxy (\citealt{lrl83}).

\subsection{4C\,73.08}

4C\,73.08 ($z=0.0581$, $d_{\rm L}=258$~Mpc) is another example of an FRII GRG. WSRT images (\citealt{may79,klein94}) show a compact core accompanied by two hotspots. The brighter (western) hotspot is connected to the core by a bridge of radio emission. Both lobes show unusual protrusions toward the north and south. Optically, 4C\,73.08 is classified as a narrow-line radio galaxy (\citealt{lrl83}).
\\ \newline
\noindent A summary of the properties of the three sources is given in Table~\ref{sourcessummary}.

\begin{table*}
\centering
\caption{Overview of the three sources}
\begin{tabular}{llcllcl}
\hline\hline
          &          &                   & Optical    & Optical spectrum & 178-MHz Luminosity Density     & Galactic \\
Source    & Redshift & FR classification & Excitation & reference        & (W Hz$^{-1}$ sr$^{-1}$) & absorption (cm$^{-2}$) \\
\hline
3C\,305   & 0.0416   & I                 & NLRG       & \cite{liu95}     & $5.50\times10^{24}$   & $1.69\times10^{20}$  \\
DA\,240   & 0.0356   & II                & LERG       & \cite{sau89}     & $5.38\times10^{24}$   & $4.36\times10^{20}$  \\
4C\,73.08 & 0.0581   & II                & NLRG       & \cite{sau89}     & $9.94\times10^{24}$   & $2.33\times10^{20}$  \\
\hline
\end{tabular}
\label{sourcessummary}
\end{table*}

\section{Observations And Data Reduction}

\begin{table*}
\centering
\caption{Observation Log}
\begin{tabular}{llllcl}
\hline\hline
Source    & Obs ID & Observation date & Filter & Nominal exposure (ks) & Screened exposure (ks) \\
\hline
3C\,305   & 0404050301 & 2006 August 08 & Medium & 11.4, 11.4, 9.8  & 11.4, 11.4, 9.8       \\
DA\,240   & 0404050101 & 2006 October 18 & Medium & 12.6, 12.6, 11.0 & 12.6, 12.6, 11.0    \\
4C\,73.08 & 0404050601 & 2007 April 28 & Medium & 4.5, 4.5, 13.9   & 3.4, 3.4, 4.7       \\
\hline
\end{tabular}
\tablecomments{Exposure times are quoted for the MOS1, MOS2, and pn cameras, respectively.}
\label{obslog}
\end{table*}

{\em XMM-Newton} observed 3C\,305, DA\,240, and 4C\,73.08 as part of AO-5. We reprocessed the data with version 7.1.0 of the Scientific Analysis Software (SAS) using the standard pipeline tasks {\sc emchain} and {\sc epchain}. The data were filtered for PATTERN values $\leq 12$ (MOS) and $\leq 4$ (pn) and the bit-mask flags 0x766a0600 (MOS) and 0xfa000c (pn). These flagsets are equivalent to the standard flagsets \#XMMEA\_EM/EP but include out of field-of-view events and exclude bad columns and rows. 

To check for intervals of high particle background, we extracted light
curves from the CCD on which the source is located. The events were
filtered to include only those with PATTERN=0 attributes and an energy
range of 10--15 keV. The background was relatively low during the
observations of 3C\,305 and DA\,240, meaning that no further filtering
was required, especially given that we are performing spectral
analyses of point sources. However, the observation of 4C\,73.08 was
heavily affected by flaring for almost the entire duration of the
observation; indeed the MOS observation was truncated due to high background. We therefore chose filtering criteria of $<2.5$~s$^{-1}$ (MOS) and $<25$~s$^{-1}$ (pn) in the 10--15 keV band to remove the worst flaring but retain sufficient data to perform spectroscopy. Table~\ref{obslog} gives the details of the three \XMM
observations. All spectral fits include Galactic absorption.

The unresolved nuclei of all three sources are detected with {\it XMM-Newton}; in addition, there is
a weak but clear detection of an X-ray source coincident with the
bright NE hotspot of DA\,240. We discuss the analysis of these X-ray
features in the following two sections.

\section{Spectroscopic Analysis of the nuclei}

\subsection{3C 305}
\label{3c305spec}

We extracted the nuclear X-ray spectrum of 3C\,305 from a source-centered circle of radius 35$''$, with background sampled from a large off-source region on the same CCD as the target. There were sufficient counts in the spectra from each of the MOS1, MOS2, and pn cameras to perform a joint analysis for all three datasets. The spectra were grouped to a minimum of 20 counts per bin.

\begin{figure}
\includegraphics[height=8cm,angle=270]{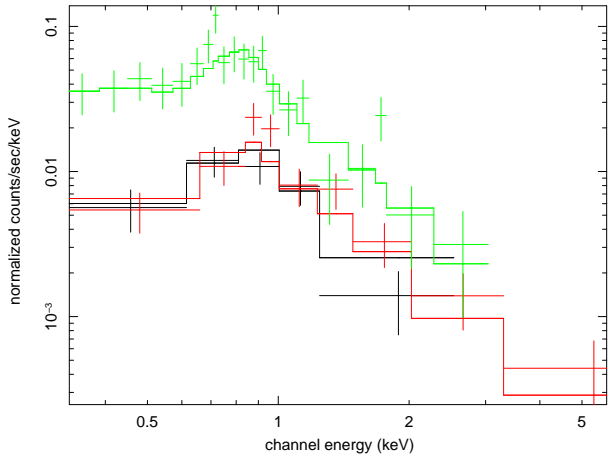}
\caption{\XMM MOS1 {\it (black)}, MOS2 {\it (red)}, and pn {\it (green)} spectrum of 3C\,305. Also shown is the best-fitting model of an unabsorbed power law and thermal emission.}
\label{3c305_spectrum}
\end{figure}

We initially attempted to fit the spectrum with a single, unabsorbed power law, but this achieved a poor fit ($\chi^{2} = 68.0$ for 33 dof). We found an acceptable fit ($\chi^{2} = 29.3$ for 31 dof) with the combination of an unabsorbed power law and thermal emission, characterized by an {\sc apec} model with $kT=0.69^{+0.10}_{-0.15}$ keV, abundance fixed at 0.3 times solar, and normalization $(3.78^{+1.06}_{-1.02})\times10^{-5}$ photons~s$^{-1}$~cm$^{-2}$~keV$^{-1}$. The best fitting parameters of the power law are $\Gamma=1.61^{+0.37}_{-0.38}$ and 1~keV normalization $(1.45\pm0.44)\times10^{-5}$ photons~s$^{-1}$~cm$^{-2}$~keV$^{-1}$. Adding additional components, such as allowing the power law to be modified by additional absorption, failed to improve the fit (the best-fitting intrinsic absorption tended to zero). The thermal interpretation is supported by a \Ch observation of the source (PI: D. Harris), which shows resolved emission elongated 3$''$ either side of the nucleus along the direction of the jets. Indeed, the unresolved \Ch nuclear flux is 3 times lower than that which we measured with {\it XMM-Newton}. The \XMM spectra and best-fitting model are shown in Figure~\ref{3c305_spectrum}.

\subsection{DA 240}

\begin{figure}
\includegraphics[height=8cm,angle=270]{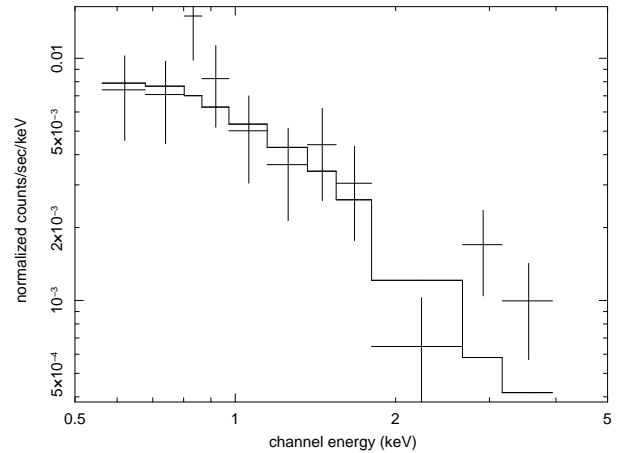}
\caption{\XMM pn spectrum of DA\,240, together with the best-fitting model of an unabsorbed power law.}
\label{da240_spectrum}
\end{figure}

We extracted the spectrum of the nucleus of DA\,240 from a source-centered circle of radius 35$''$, and extracted a background spectrum from a large off-source circular region on the same CCD. Only the spectrum from the pn camera had sufficient counts for an adequate spectral analysis. With the data grouped to 10 counts per bin, we found an acceptable fit ($\chi^{2} = 10.5$ for 9 dof) with a single unabsorbed power law of photon index $1.91^{+0.54}_{-0.51}$ and 1~keV normalization $(6.48^{+1.48}_{-1.51})\times10^{-6}$ photons~s$^{-1}$~cm$^{-2}$~keV$^{-1}$. Allowing the power law to be modified by intrinsic absorption failed to improve the fit (the best-fitting $N_{\rm H}$ is zero). Additional components to our model also led to no statistically significant improvement in the fit.  The spectrum, and best-fitting model, of a single, unabsorbed power law, are shown in Figure~\ref{da240_spectrum}.

\subsection{4C 73.08}

\begin{figure}
\includegraphics[height=8cm,angle=270]{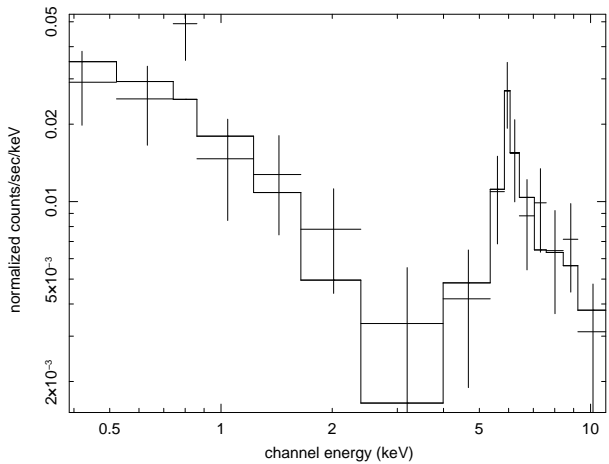}
\caption{\XMM pn spectrum of 4C\,73.08. Also shown is the best-fitting model of a heavily absorbed power law, a neutral Fe K$\alpha$ line, and a second, unabsorbed power law.}
\label{4c73.08_spectrum}
\end{figure}

We sampled the nuclear spectrum of 4C\,73.08 from a source-centered circle of radius 35$''$, and extracted a background spectrum from a large off-source circular region on the same chip. We initially attempted to model the spectrum with a single, unabsorbed power law, but this achieved a poor fit ($\chi^{2} = 63.2$ for 15 dof), and we noticed significant residuals above $\sim$4~keV that clearly indicated the presence of an additional, heavily absorbed component. We achieved a good fit ($\chi^{2} = 7.2$ for 11 dof) to the spectrum with the sum of a heavily absorbed [$N_{\rm H}=(9.2^{+5.4}_{-2.9})\times10^{23}$ cm$^{-2}$] power law of photon index frozen at 1.7, a Gaussian neutral, unresolved, Fe K$\alpha$ line of equivalent width $\sim$300~eV (the Gaussian line is significant at the 2$\sigma$ level), and a second, unabsorbed, power law of photon index frozen at 2. There are insufficient counts to fit the power-law slopes, so we adopted values consistent with canonical values found in radio galaxies (e.g., \citealt{evans04,evans06}). The 1~keV normalizations of the power laws are $(1.82^{+2.19}_{-1.00})\times10^{-3}$  photons~s$^{-1}$~cm$^{-2}$~keV$^{-1}$ and $(1.90\pm0.47)\times10^{-5}$  photons~s$^{-1}$~cm$^{-2}$~keV$^{-1}$, respectively. Replacing the unabsorbed power law with a thermal component resulted in a worse fit to the spectrum ($\chi^{2} = 16.9$ for 10 dof). The spectrum and our best-fitting model are shown in Figure~\ref{4c73.08_spectrum}. The hotspots of 4C\,73.08 are not detected in this short observation. \\

\noindent 
Table~\ref{spectralfitting} summarizes the best-fitting spectral models to the X-ray spectra of 3C\,305, DA\,240, and 4C\,73.08.

\begin{table*}
\centering
\caption{Summary of best-fitting spectra}
\begin{tabular}{lllllllll}
\hline\hline
           &                                &                          &             &            &                 &      & $L_{\rm (2-10 keV)}$                         &                    \\
           &                                &                          &             &            &                 &      & (Power Law)                                  &                    \\
    Source &   Spectrum                     &  $N_{\rm H}$ (cm$^{-2}$) &   $\Gamma$  &    E (keV) &  $\sigma$ (keV) & $kT$ (keV) & (ergs s$^{-1}$) &  $\chi^{2}$/dof  \\
    (1)    &             (2)                &   (3)                    &     (4)     &     (5)    &      (6)        &      (7)   &             (8) &          (9)     \\
\hline
    3C 305 & PL+TH &         --             & $1.61^{+0.37}_{-0.38}$   &         --  &         -- & $0.61^{+0.10}_{-0.15}$ & $(2.6\pm0.8)\times10^{41}$ & 29.3/31           \\
    DA 240 &    PL &         --             & $1.91^{+0.54}_{-0.51}$   &         --  &         -- &         --             & $(5.5\pm1.3)\times10^{40}$         & 10.5/9            \\
  4C 73.08 & $N_{\rm H}$(PL+Gauss)+PL & $(9.2^{+5.4}_{-2.9})\times10^{23}$ & $\Gamma_1 = 1.7$ (f);& 6.32 $\pm$ 0.14 &    0.01 (f) &         -- & $(5.7^{+6.9}_{-3.2})\times10^{43}$ & 7.2/11           \\
           &                &                    --                 & $\Gamma_2 = 2$ (f)  &      --         &     --     &          --        & $(3.8\pm0.9)\times10^{41}$ &                 \\
\hline
\end{tabular}
\tablecomments{Col. (1): Name of source. Col. (2): Description of best spectrum ($N_{\rm H}$=Intrinsic absorption, PL=Power Law, Gauss=Redshifted Gaussian Line, TH=Thermal. Col. (3): Intrinsic neutral hydrogen column density. Galactic absorption has also been applied. Col. (4): Power-law photon index. Col. (5): Rest-frame Gaussian centroid energy. Col. (6): Gaussian linewidth. Col. (7): Temperature of thermal component. Col. (8): 2--10 keV unabsorbed luminosity of primary power law. Col. (9): Value of $\chi^{2}$ and degrees of freedom. (f) Indicates parameter was frozen.}
\label{spectralfitting}
\end{table*}

\begin{table*}
\centering
\caption{X-ray and radio luminosities}
\begin{tabular}{llllll}
\hline\hline
& Radio core luminosity & 1-keV soft luminosity & 178-MHz luminosity & 2--10 keV `accretion' & Core luminosity \\ 
Source & density (W Hz$^{-1}$ sr$^{-1}$) and frequency & density (W Hz$^{-1}$ sr$^{-1}$) & density (W Hz$^{-1}$ sr$^{-1}$) & luminosity (ergs s$^{-1}$) & Reference \\ 
(1) & (2) & (3) & (4) & (5) & (6) \\
\hline
3C 305  & $<$$3.21\times10^{20}$ (1.4 GHz) & $(3.1\pm0.9)\times10^{15}$ & $5.50\times10^{24}$ & $<1.0\times10^{42}$ & \cite{sar97} \\ 
DA 240  & $2.46\times10^{22}$    (5 GHz)   & $(1.4\pm0.3)\times10^{15}$ & $5.38\times10^{24}$ & $<5.3\times10^{41}$ & \cite{tsi82} \\ 
4C73.08 & $3.53\times10^{21}$    (5 GHz)   & $(4.1\pm1.0)\times10^{15}$ & $9.94\times10^{24}$ & $(5.7^{+6.9}_{-3.2})\times10^{43}$ & \cite{sar97} \\ 
\hline
\end{tabular}
\tablecomments{Col. (1): Name of source. Col. (2): Radio luminosity density of core. Col. (3): 1-keV unabsorbed luminosity density of soft X-ray component. Col. (4): 178-MHz VLA luminosity density. Col. (5): 2--10 keV unabsorbed X-ray `accretion-related' luminosity. Obtained by direct fitting with free $N_{\rm H}$ (4C\,73.08) and as a hidden component with $N_{\rm H}$ fixed at $10^{23}$ cm$^{-2}$ (3C\,305, DA\,240). Col. (6): Reference for radio core luminosity density.}
\label{luminosities}
\end{table*}

\section{The Hotspot Of DA\,240}

\begin{figure}
\includegraphics[width=8.5cm,angle=0]{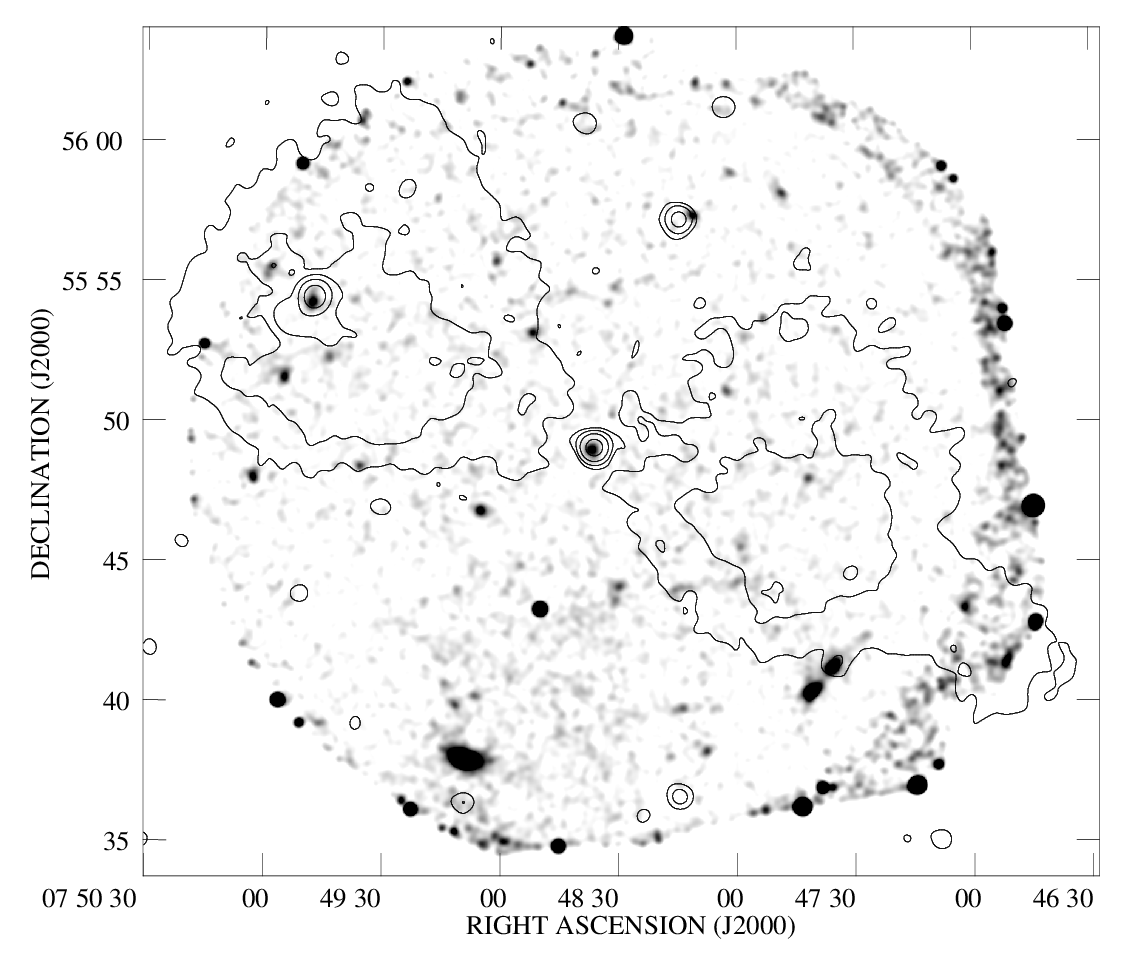}
\caption{\XMM observations of DA\,240 in the MOS and pn cameras, co-added taking
account of exposure and smoothed with a Gaussian of FWHM $15.3''$.
Overlaid are contours from the $34''$-resolution 608-MHz WSRT map
described in the text, at $2 \times(1,4,16\dots)$ mJy beam$^{-1}$.}
\label{da240-xmm}
\end{figure}

The bright NE hotspot of the giant radio galaxy DA\,240 lies in the \XMM
field of view and is clearly detected in X-rays in our observation
(Fig.\ \ref{da240-xmm}). X-ray detections of the hotspots in
relative low-luminosity radio sources like our targets are quite common in {\it
Chandra} observations \citep[e.g.,][]{khwm05,hc05,kbhe07,hck07,efhk08} and recently some bright hotspots have also been
detected with \XMM \citep{efbm07,ghck08}.
It has been argued that that X-ray detections of
low-luminosity hotspots such as those of DA\,240 (whose NE hotspot has a 5-GHz radio luminosity of $9\times10^{22}$ W~Hz$^{-1}$~sr$^{-1}$) are almost certainly due to
synchrotron rather than inverse-Compton emission (see Fig.~5 of \citealt{hhwb04}). If this is the case,
X-ray detections of hotspots can give us important information about
the relationship between the location of high-energy particle
acceleration (traced by the X-ray) and the locations where low-energy
particle and field energy densities are highest (traced by the radio
synchrotron emission). The available evidence to date is that this
relationship is complex; kpc-scale offsets are often found between the
peaks of X-ray and radio emission \citep[e.g.,][]{hck07}.

\begin{figure}
\includegraphics[width=8.5cm,angle=0]{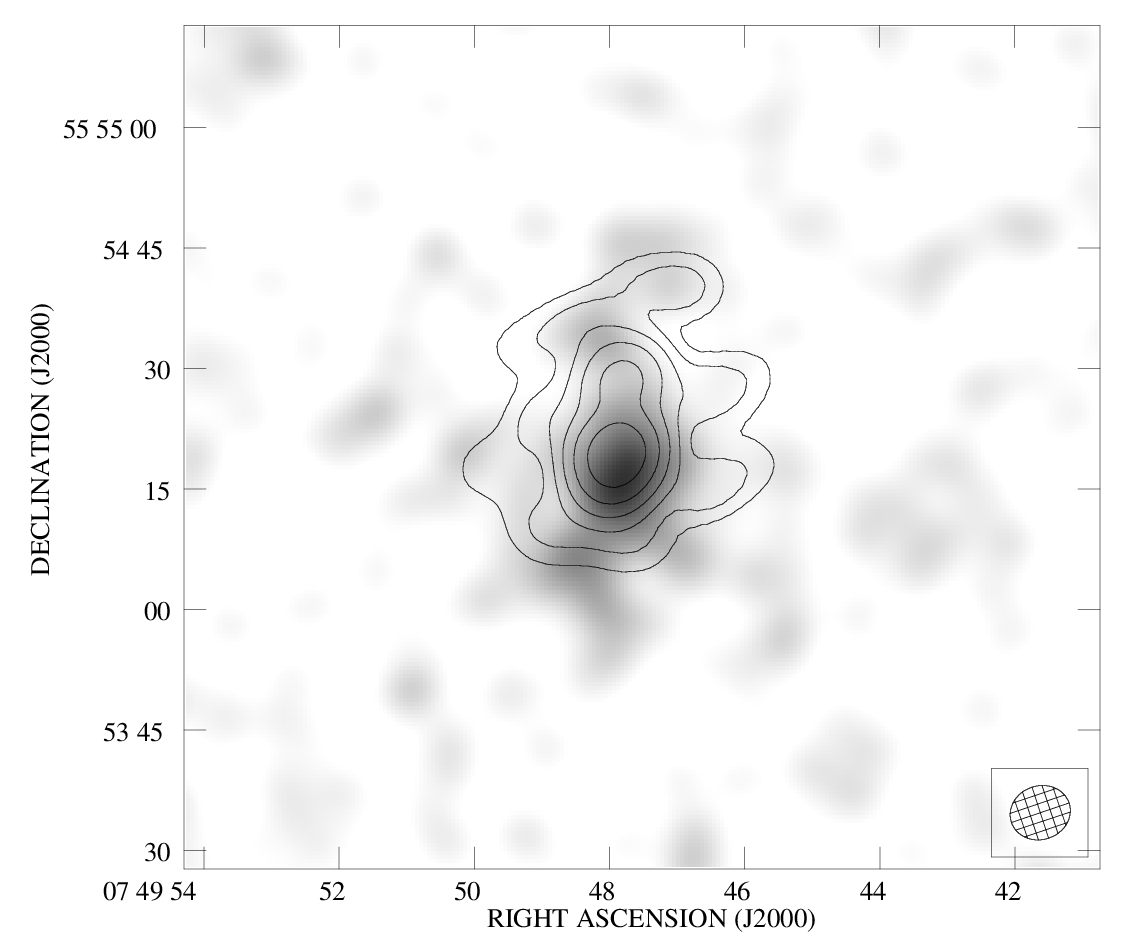}
\caption{Co-added \XMM observations of DA\,240 as in Fig.\
  \ref{da240-xmm}, but smoothed with a Gaussian with FWHM $6.1''$.
  Overlaid are contours of our VLA data with $7.5'' \times 6.7''$
  resolution (major $\times$ minor axis of restoring elliptical
  Gaussian) at $4 \times (1, 2, 4\dots)$ mJy beam$^{-1}$. The 90\% off-axis encircled energy radius of \XMM at this location is $\sim$1$'$.}
\label{da240-zoom}
\end{figure}

The only radio image for DA\,240 available to us at the start of our
study was the WSRT 608-MHz image from the on-line atlas of
low-$z$ 3CRR sources\footnote{http://www.jb.man.ac.uk/atlas/}. This
image has a resolution of $34''$, and so does not allow us to see
details of the hotspot structure or its relationship to the X-ray
emission. Accordingly we obtained a short observation of the hotspot
with the VLA at 4.9 GHz under the exploratory time program. This
observation (observation identifier AE163) was taken on 2007 May 18
when the VLA was in the process of moving between its D and A
configurations. In addition, the EVLA antennas of the array were
unavailable for most of the observation. As a result there were only
13 antennas available in the expected D-configuration arrangement, as
opposed to the usual 27. Nevertheless we obtained an image with a
resolution of $7.5''$ and were able to detect and resolve the compact
hotspot (Fig.\ \ref{da240-zoom}). The radio emission coincident with the
X-ray emission is resolved into two compact components aligned roughly
N-S, with the brighter southern component having a radio flux density
of 270 mJy and the fainter northern component at a level of $\sim 90$
mJy, plus extended structure. Both the compact components are point-like at
the resolution of our image; however, higher-resolution images (\citealt{tsi82}) resolve the southern component, showing it contains at least two separate peaks. The peak of the X-ray emission is closest
to the brighter component of the hotspot, but both components may be
X-ray sources: the resolution of \XMM is not good enough (particularly
at this off-axis distance) to separate them. We used the default astrometry for both the VLA and \XMM data to search for offsets between the radio and X-ray emission in the northern hotspot (the radio core is too far down the primary beam of the VLA for us to be able to use it to align the radio and X-ray frames). The X-ray emission appears to be offset by several kpc approximately in the direction of the nucleus. However the higher-resolution radio observation of \cite{tsi82}, which shows both the nucleus and hotspot, indicates that the brightest radio hotspot subcomponent (marked A by Tsien et al.) is separated by approximately 8$''$ ($\sim$ 5.5~kpc) from the brightest X-ray hotspot emission. This is consistent with our VLA observation in both sense and approximate offset with respect to the X-ray emission. Offsets this large, or larger, have been observed in other powerful radio galaxies (e.g., \citealt{efbm07}).

We extracted a spectrum for the hotspot from
the pn data, using a 30$''$ radius circle as the source region and
local background, and fitted it with a power-law model with Galactic
absorption. It is well fitted ($\chi^2 = 0.29$ for 2 d.o.f.) with a
model with photon index of $2.2 \pm 0.3$ and 1-keV unabsorbed flux density of $7 \pm
1$ nJy. This flux density puts it among the brighter known X-ray
hotspots \citep{hhwb04}, very similar in flux density
(though not luminosity) to the bright hotspot detected by \citet{efbm07} in the giant quasar 4C\,74.26.

The steep X-ray spectrum and the possible offset
between the radio and X-ray peak favor a synchrotron rather than
inverse-Compton origin for the X-rays in this source. In the absence of optical
measurements it is easy to fit a curved or broken synchrotron spectrum
through the radio and X-ray data. Moreover, if we
model the hotspot (normalizing using the observed radio flux for the
brighter component) as a uniform sphere at equipartition with a radius
of 1 kpc (which is consistent with the size reported by \citealt{tsi82} for
 the most compact component of the hotspot only) then the predicted inverse-Compton flux density, using the code
of \citet{hbw98}, is 3 orders of magnitude below that observed. All
of the flux density from the hotspot would have to come from a region
$<0.1$ pc in size, the magnetic field strength would have to be a
factor $\sim 30$ below the equipartition value, or some combination of
the two would have to apply in order for the observed X-ray flux
density to be produced by the synchrotron self-Compton model. Since
DA240 is a low-excitation radio galaxy, the nuclear emission-line
classification gives us no information about the orientation with
respect to the line of sight, and so it is possible in principle that
inverse-Compton emission could be boosted by a process which requires
beaming and small angles of the jet to the line of sight
\citep[e.g.,][]{gk03}. However, an efficient role for beaming would
imply a very large physical size ($\ga 4$ Mpc) for DA240, so we do not
regard this model as probable, and we attribute the X-ray emission to synchrotron radiation. The relative brightness
of the hotspot should make it a good target for follow-up radio and
high-resolution X-ray observations aimed at understanding the details of high-energy
particle acceleration in this source.

\section{Interpretation of the nuclear spectra}
\label{interp}

\begin{figure}
\includegraphics[width=8cm,angle=0]{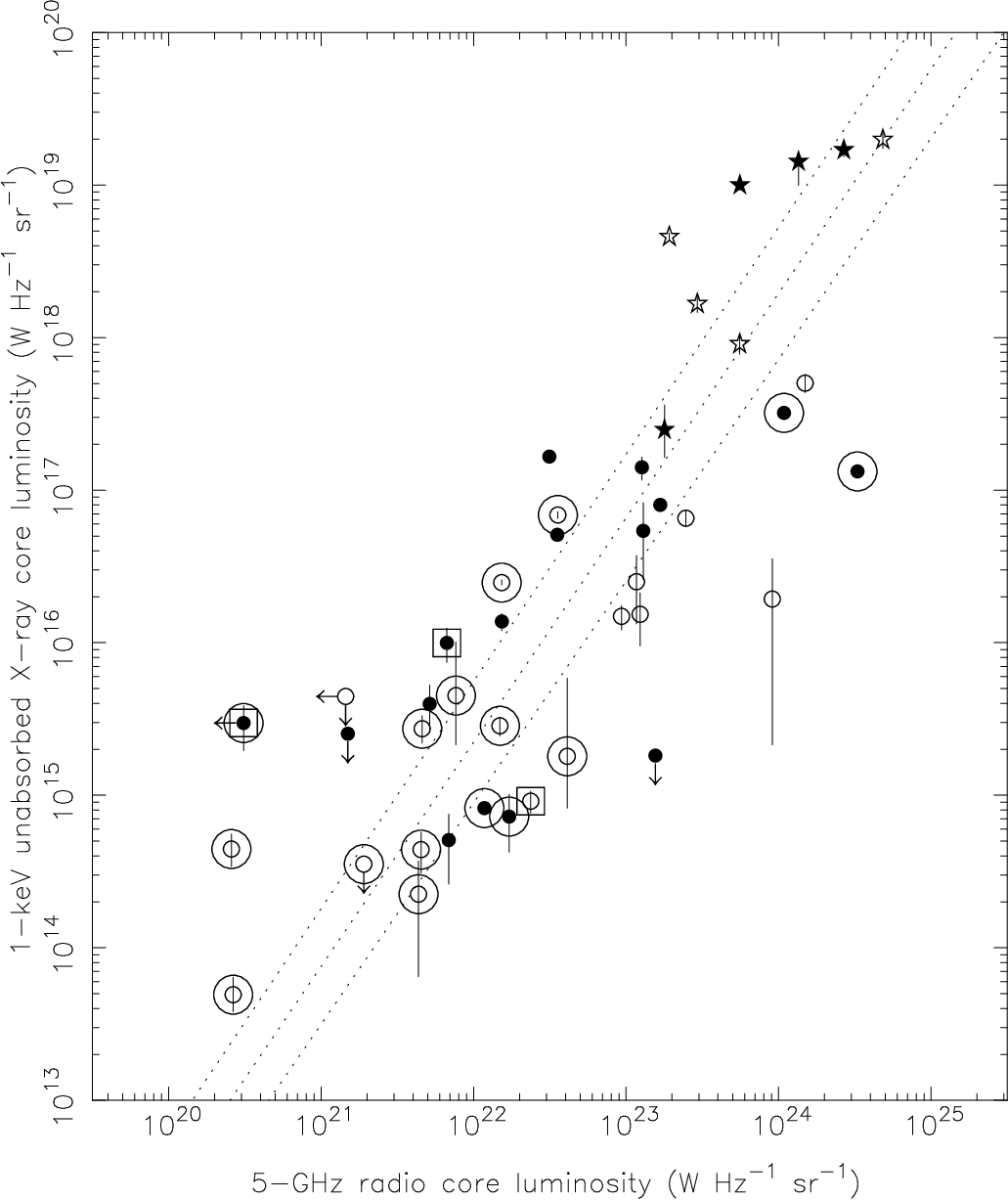}
\caption{X-ray luminosity of the unabsorbed nuclear component for the three sources, together with combined $z<0.5$ sample (\citealt{evans06,hec06}) as a function of 5-GHz radio core luminosity (Table \ref{luminosities}). Open circles are LERG, filled circles NLRG, open stars BLRG, and filled stars quasars. Large surrounding circles indicate that a source is an FRI. The sources studied in this paper are indicated by surrounding boxes. Note that the core luminosity of 3C\,305 is measured at 1.4 GHz. Where error bars are not visible they are smaller than symbols. Dotted lines show the regression line to all data and its 1$\sigma$ confidence range.}
\label{rx}
\end{figure}

\begin{figure}
\includegraphics[width=8cm,angle=0]{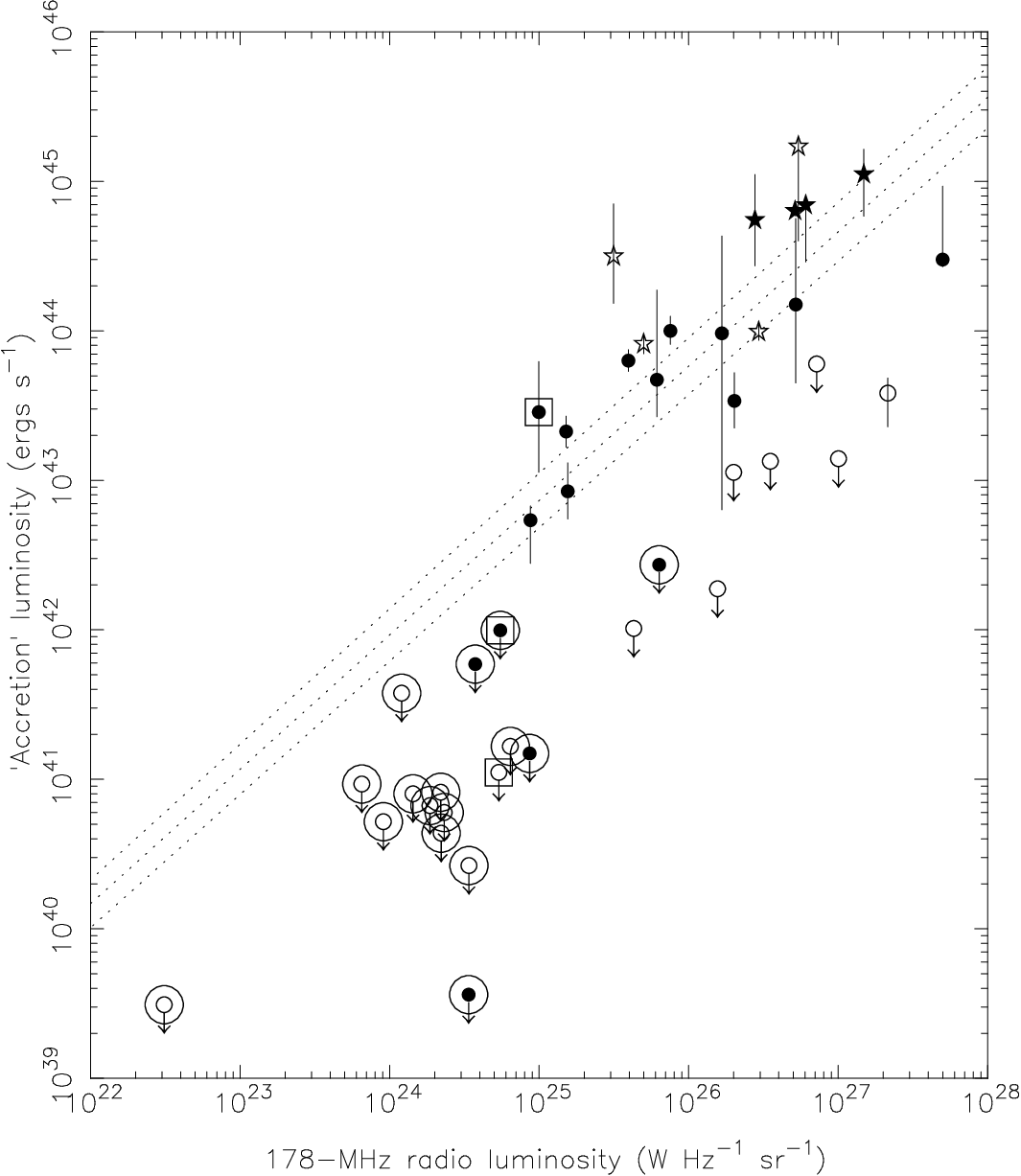}
\caption{X-ray luminosity of the accretion-related component for the three sources, together with combined $z<0.5$ sample (\citealt{evans06,hec06}) as a function of 178-MHz total radio luminosity (Table \ref{luminosities}). Symbols are as in Fig.\ \ref{rx} .Dotted lines show the regression line to the NLRGs only and its 1$\sigma$ confidence range.}
\label{lum-lum}
\end{figure}

\subsection{Overview of the spectra}
\label{interp_spectra}

The unabsorbed X-ray spectrum of the low-excitation FRII radio galaxy DA\,240 is consistent with the other LERGs in the \cite{hec06} sample, which observationally encompass both FRI and FRII radio galaxies. The high-excitation (narrow-line) FRII radio galaxy 4C\,73.08 shows a heavily absorbed, luminous, component of X-ray emission, similar to the other narrow-line radio galaxies studied by \cite{hec06}. However, the spectrum of 3C\,305 shows no evidence for the heavily absorbed X-ray emission that is characteristic of narrow-line radio galaxies. We return to this in Section~\ref{interp_nlrgs}.

\subsection{The Radio Core--X-Ray Core Correlation}

Figure~\ref{rx} shows a plot of the 1-keV luminosity of the low-absorption power-law component against the core luminosity for our three sources (Table~\ref{luminosities}), together with the 3CRR sources presented in \cite{evans06} and \cite{hec06}. DA\,240 and 4C\,73.08 both lie close to the correlation between the radio and X-ray luminosities established by, e.g., \cite{fab84}, \cite{wb94}, and \cite{hec06}. On the other hand, 3C\,305 lies somewhat away from the other data, though this is almost certainly due to the extended X-ray emission detected with {\it Chandra}. Indeed, the unresolved \Ch nuclear flux is 3 times lower than the \XMM value, which would bring 3C\,305 closer to established trendline in Figure~\ref{rx}. The radio--X-ray core correlation suggests a common origin of the two at the base of an unresolved jet, as has been extensively argued by, e.g., \cite{wb94}, \cite{har99}, \cite{evans06}, \cite{bal06} and \cite{bel06}.

\subsection{Accretion-Related X-ray Emission}

We now wish to consider any accretion-related X-ray emission. For 4C\,73.08, as with the other narrow-line radio galaxies studied by \cite{evans06} and \cite{hec06}, we can take the accretion-related luminosity to be the unobscured luminosity of the heavily absorbed power-law component: this is supported by the presence of Fe K$\alpha$ emission (e.g., \citealt{evans06}). The 2--10 keV accretion luminosity of 4C\,73.08 ($\sim$6$\times10^{43}$ ergs s$^{-1}$) is substantially larger than its jet-related luminosity ($\sim$4$\times10^{41}$ ergs s$^{-1}$), as has been found with other NLRGs (e.g., \citealt{evans06,hec06}).

In the cases of the LERG DA\,240 and the (purported NLRG 3C\,305), we followed the method of \cite{evans06} and assumed that, in addition to the dominant jet component of X-ray emission, there exists an additional `hidden' component of accretion-related emission of photon index 1.7 that is obscured by a torus of intrinsic absorption $10^{23}$ cm$^{-2}$. We added this component to the best-fitting model, refitted the spectra, and determined the 90\%-confidence upper limit to the 2--10 keV accretion-related luminosity to be $5.3\times10^{41}$ ergs~s$^{-1}$ for DA\,240 and $1.0\times10^{42}$ ergs~s$^{-1}$ for 3C\,305.

Figure~\ref{lum-lum} shows a plot of the 178-MHz and 2--10 keV accretion-related luminosities of the three sources (Table~\ref{luminosities}), together with those of the other $z<0.5$ 3CRR sources studied by \cite{evans06} and \cite{hec06}. Figure~\ref{lum-lum} shows that the upper limit to the accretion-related components in the LERG DA\,240, given our assumed absorbing column of $10^{23}$ cm$^{-2}$, lies below the trendline established for high-excitation (narrow-line) radio galaxies such as 4C\,73.08. Of course, if no obscuring region is present in DA\,240, as seems to be the case in other LERGs, then the luminosity of any accretion-related emission will be substantially lower than that shown. Alternatively, the accretion-related X-ray luminosity of LERGs can be made to lie in the region occupied by the HERGs, but this requires extremely high values of intrinsic absorption (\citealt{evans04}) that can be ruled out by infrared observations (e.g, \citealt{mul04}). The upper limit to the accretion-related luminosity of 3C\,305 lies between the populations of low- and high-excitation sources in Figure~\ref{lum-lum}, though as previously mentioned the {\it XMM-Newton}-measured unresolved core flux is overestimated by a factor $\sim$3 (see Section~\ref{3c305spec}).

\subsection{Optical Emission Line Classifications and Relationships to the Central Engine}
\label{interp_nlrgs}

In our previous studies of the X-ray properties of 3CRR radio sources we have used the \cite{lrl83} optical emission-line classifications of low- and high-excitation radio galaxies. \cite{laing94} provided a quantitative definition of LERGs as having [OIII] equivalent widths of less than 3~\AA\ and [OIII]/H$\alpha$ line ratios $>0.2$. A similar classification was given by \cite{jac97}, with LERGs having [OIII] equivalent widths of less than 10~\AA\ and/or [OII]/[OIII] line ratios $>1$. However, these definitions of low- and high-excitation sources do not necessarily take into account the potentially {\it different} sources of ionizing radiation, their size scale, or their relationship to the AGN itself. This may lead to occasional ambiguities where sources are classified based on their emission-line characteristics. We discuss some of the issues here.

{\it HST} observations of the nuclei of radio galaxies have revealed the origin of the optical continuum emission and its likely relationship to any unresolved emission lines. In the case of LERGs, \cite{chi99} and \cite{har00} showed that the correlations between the radio and optical continuum luminosities support the common origin of the two in the form of a jet.{\it HST} narrow-band imaging of LERGs (\citealt{cap05}) showed that so-called Compact Emission Line Regions (CELRs) are commonplace, and that they are associated with the dominant source of ionizing photons, assumed to be the jet. Further, \cite{chi02} argued that the dominant contribution to the optical emission in obscured high-excitation radio galaxies is the accretion disk. There is likely to be a substantial ionizing field in these sources that is directly related to the accretion process.

On larger scales, high-resolution {\it HST} emission-line images of the extended environments in radio galaxies (\citealt{pri08}) provide insights on the different components that constitute the kpc-scale narrow-line region (NLR) and $\sim$10~kpc-scale extended narrow-line region (ENLR). In addition to photoionization from the nucleus, jet--environment interactions may play a significant role in governing the energy budget of the NLR and ENLR, either in the form of collisional ionization or a radiative 'autoionizing' shock (e.g., \citealt{ds95,ds96}).

The different physical origins for optical line-emission in radio
galaxies illustrate the difficulties in disentangling genuine AGN
emission from that which is not directly related to the accretion
process. An excellent case in point is the purported NLRG 3C\,305,
whose X-ray spectrum is consistent with that of a LERG, rather than a
NLRG. {\it HST} [OII] observations of the extended
emission-line environment in the source show that the majority of the
[OII] emission lies just beyond the edge of the radio jet at a
distance of 1.5$''$ from the core, and \cite{jac95} suggested that it
has been shock-excited by the jet. Several other FRI radio sources studied
by \cite{evans06} and \cite{hec06} also show optical spectra that may be
attributed to their environments. Some of these lie at the centers of
cooling-core clusters, in which significant amounts of optical
line-emission might be expected that are not necessarily directly related to the central AGN. This may go some way to explaining the handful of other purported NLRGs in Figure~\ref{lum-lum} whose X-ray properties are more consistent with low-excitation sources.

The above arguments suggest that the emission-line classification of relatively weak-lined radio galaxies does not always reflect the nuclear accretion activity itself. We propose that only the combination of high-resolution optical spectroscopy, X-ray observations, and constraints from {\it Spitzer} mid-infrared observations of reprocessed emission in radio galaxies can reliably determine the structure of the central engine in radio-loud AGN. In the case of 3C\,305, {\it Spitzer} observations would enable us to distinguish between (1) a genuinely narrow-line radio galaxy that is obscured by a Compton-thick absorber (in which case the $<10$~keV X-ray continuum would show few, if any, signs of heavily absorbed emission), and (2) a low-excitation radio galaxy with a prominent extended emission-line environment. We will return to this point in subsequent publications (Birkinshaw et al., 2008, in prep.; Hardcastle et al. 2008, in prep.).

\section{Conclusions}

We have presented results from \XMM observations of the nuclei of the radio galaxies 3C\,305, DA\,240, and 4C\,73.08. We have shown the following:

\begin{enumerate}
\item The X-ray spectrum of the narrow-line FRII radio galaxy 4C\,73.08 can be modeled as the sum of a heavily absorbed power law associated with a luminous accretion disk and circumnuclear obscuring structure, together with an unabsorbed component of X-ray emission that has a common origin with the radio emission at the base of an unresolved jet. This behavior is consistent with the other narrow-line FRII radio galaxies studied by \cite{evans06} and \cite{hec06}.
\item The nuclear X-ray spectrum of the FRII giant radio galaxy DA\,240, optically classified as a low-excitation radio galaxy, can be modeled as a single, unabsorbed power law that is likely associated with emission from the parsec-scale jet. The upper limit to the X-ray luminosity of any additional, accretion-related emission suggests that the accretion process in DA\,240 is substantially sub-Eddington and likely radiatively inefficient in nature.
\item The X-ray emission in the nucleus of the narrow-line radio galaxy 3C\,305 can be modeled as an unabsorbed power law that originates at the base of the jet. However, it shows no evidence for heavily absorbed X-ray emission was found in the NLRGs studied by \cite{evans06}.
\item We have discovered an X-ray counterpart to the NE hotspot of the
  giant radio galaxy DA\,240. We argue that the emission process is
  overwhelmingly likely to be synchrotron emission. Because of the
  high X-ray flux of the hotspot, it is a good candidate for followup
  high-resolution X-ray observations.
\item We have discussed the different origins of optical emission lines in the nuclear and circumnuclear gaseous environments of radio galaxies. These include photoionization from the AGN accretion flow or parsec-scale jet, shock-excitation by the radio jet, or cooling gas in the centers of clusters. This may lead to occasional misclassification of genuinely weak-lined sources such as 3C\,305 as high-excitation sources.
\item We therefore argue that there is not necessarily always a one-to-one correspondence between optical emission-line class (low- vs. high-excitation) and accretion-flow state (inefficient flow vs. standard thin disk), especially when low angular-resolution optical spectroscopy is used. We suggest that only the combination of high-resolution optical, X-ray, and infrared observations can reliably uncover the nature of the central engine in radio-loud AGN.
\end{enumerate}

\acknowledgements

DAE gratefully acknowledges financial support for this work from NASA
under grant number NNX06AG37G. MJH thanks the Royal Society for a
Research Fellowship. We wish to thank the anonymous referee for valuable 
comments. We also thank Dan Harris and Francesco Massaro for useful 
discussions of the nuclear properties of 3C\,305. This work is based 
on observations obtained with
{\it XMM-Newton}, an ESA science mission with instruments and
contributions directly funded by ESA Member States and NASA. The
National Radio Astronomy Observatory is a facility of the National
Science Foundation operated under cooperative agreement by Associated
Universities, Inc. This research has made use of the NASA/IPAC
Extragalactic Database (NED) which is operated by the Jet Propulsion
Laboratory, California Institute of Technology, under contract with
the National Aeronautics and Space Administration.

\end{document}